\documentclass[twocolumn,letterpaper,aps,prd,longbibliography,superscriptaddress,nofootinbib,floatfix]{revtex4-1}
\usepackage[normalem]{ulem}
\usepackage{graphicx}	
\usepackage{natbib}
\usepackage{amsmath}
\usepackage{pifont}
\usepackage{multirow}
\usepackage{textcomp}
\usepackage{xcolor}

\usepackage{url}

\usepackage{xspace}	
\usepackage{mathtools}

\newcommand{\pt}{\mbox{$p_T$}\xspace}

\newcommand{\bb}{$b\bar{b}$\xspace}
\newcommand{\pp}{$p$+$p$\xspace}
\newcommand{\dphi}{$\Delta\phi$\xspace}
\newcommand{\accEff}{$A\epsilon$\xspace}

\begin{document}

\title{Production of $b\bar{b}$ at forward rapidity in $p$+$p$ 
collisions at $\sqrt{s}=510$ GeV}

\newcommand{\abilene}{Abilene Christian University, Abilene, Texas 79699, USA}
\newcommand{\augie}{Department of Physics, Augustana University, Sioux Falls, South Dakota 57197, USA}
\newcommand{\banaras}{Department of Physics, Banaras Hindu University, Varanasi 221005, India}
\newcommand{\barc}{Bhabha Atomic Research Centre, Bombay 400 085, India}
\newcommand{\baruch}{Baruch College, City University of New York, New York, New York, 10010 USA}
\newcommand{\bnlcoll}{Collider-Accelerator Department, Brookhaven National Laboratory, Upton, New York 11973-5000, USA}
\newcommand{\bnlphys}{Physics Department, Brookhaven National Laboratory, Upton, New York 11973-5000, USA}
\newcommand{\caucr}{University of California-Riverside, Riverside, California 92521, USA}
\newcommand{\charlesczech}{Charles University, Ovocn\'{y} trh 5, Praha 1, 116 36, Prague, Czech Republic}
\newcommand{\cns}{Center for Nuclear Study, Graduate School of Science, University of Tokyo, 7-3-1 Hongo, Bunkyo, Tokyo 113-0033, Japan}
\newcommand{\colorado}{University of Colorado, Boulder, Colorado 80309, USA}
\newcommand{\columbia}{Columbia University, New York, New York 10027 and Nevis Laboratories, Irvington, New York 10533, USA}
\newcommand{\czechtech}{Czech Technical University, Zikova 4, 166 36 Prague 6, Czech Republic}
\newcommand{\debrecen}{Debrecen University, H-4010 Debrecen, Egyetem t{\'e}r 1, Hungary}
\newcommand{\elte}{ELTE, E{\"o}tv{\"o}s Lor{\'a}nd University, H-1117 Budapest, P{\'a}zm{\'a}ny P.~s.~1/A, Hungary}
\newcommand{\eszterhazy}{Eszterh\'azy K\'aroly University, K\'aroly R\'obert Campus, H-3200 Gy\"ongy\"os, M\'atrai \'ut 36, Hungary}
\newcommand{\ewha}{Ewha Womans University, Seoul 120-750, Korea}
\newcommand{\famu}{Florida A\&M University, Tallahassee, FL 32307, USA}
\newcommand{\fsu}{Florida State University, Tallahassee, Florida 32306, USA}
\newcommand{\gsu}{Georgia State University, Atlanta, Georgia 30303, USA}
\newcommand{\hanyang}{Hanyang University, Seoul 133-792, Korea}
\newcommand{\hiroshima}{Hiroshima University, Kagamiyama, Higashi-Hiroshima 739-8526, Japan}
\newcommand{\howard}{Department of Physics and Astronomy, Howard University, Washington, DC 20059, USA}
\newcommand{\ihepprot}{IHEP Protvino, State Research Center of Russian Federation, Institute for High Energy Physics, Protvino, 142281, Russia}
\newcommand{\illuiuc}{University of Illinois at Urbana-Champaign, Urbana, Illinois 61801, USA}
\newcommand{\inrras}{Institute for Nuclear Research of the Russian Academy of Sciences, prospekt 60-letiya Oktyabrya 7a, Moscow 117312, Russia}
\newcommand{\instpasczech}{Institute of Physics, Academy of Sciences of the Czech Republic, Na Slovance 2, 182 21 Prague 8, Czech Republic}
\newcommand{\isu}{Iowa State University, Ames, Iowa 50011, USA}
\newcommand{\jaea}{Advanced Science Research Center, Japan Atomic Energy Agency, 2-4 Shirakata Shirane, Tokai-mura, Naka-gun, Ibaraki-ken 319-1195, Japan}
\newcommand{\jeonbuk}{Jeonbuk National University, Jeonju, 54896, Korea}
\newcommand{\jyvaskyla}{Helsinki Institute of Physics and University of Jyv{\"a}skyl{\"a}, P.O.Box 35, FI-40014 Jyv{\"a}skyl{\"a}, Finland}
\newcommand{\kek}{KEK, High Energy Accelerator Research Organization, Tsukuba, Ibaraki 305-0801, Japan}
\newcommand{\korea}{Korea University, Seoul 02841, Korea}
\newcommand{\kurchatov}{National Research Center ``Kurchatov Institute", Moscow, 123098 Russia}
\newcommand{\kyoto}{Kyoto University, Kyoto 606-8502, Japan}
\newcommand{\lahorelums}{Physics Department, Lahore University of Management Sciences, Lahore 54792, Pakistan}
\newcommand{\lawllnl}{Lawrence Livermore National Laboratory, Livermore, California 94550, USA}
\newcommand{\losalamos}{Los Alamos National Laboratory, Los Alamos, New Mexico 87545, USA}
\newcommand{\lund}{Department of Physics, Lund University, Box 118, SE-221 00 Lund, Sweden}
\newcommand{\maryland}{University of Maryland, College Park, Maryland 20742, USA}
\newcommand{\mass}{Department of Physics, University of Massachusetts, Amherst, Massachusetts 01003-9337, USA}
\newcommand{\michigan}{Department of Physics, University of Michigan, Ann Arbor, Michigan 48109-1040, USA}
\newcommand{\muhlenberg}{Muhlenberg College, Allentown, Pennsylvania 18104-5586, USA}
\newcommand{\myongji}{Myongji University, Yongin, Kyonggido 449-728, Korea}
\newcommand{\nara}{Nara Women's University, Kita-uoya Nishi-machi Nara 630-8506, Japan}
\newcommand{\natmephi}{National Research Nuclear University, MEPhI, Moscow Engineering Physics Institute, Moscow, 115409, Russia}
\newcommand{\newmex}{University of New Mexico, Albuquerque, New Mexico 87131, USA}
\newcommand{\nmsu}{New Mexico State University, Las Cruces, New Mexico 88003, USA}
\newcommand{\northcg}{Physics and Astronomy Department, University of North Carolina at Greensboro, Greensboro, North Carolina 27412, USA}
\newcommand{\ohio}{Department of Physics and Astronomy, Ohio University, Athens, Ohio 45701, USA}
\newcommand{\ornl}{Oak Ridge National Laboratory, Oak Ridge, Tennessee 37831, USA}
\newcommand{\orsay}{IPN-Orsay, Univ.~Paris-Sud, CNRS/IN2P3, Universit\'e Paris-Saclay, BP1, F-91406, Orsay, France}
\newcommand{\pnpi}{PNPI, Petersburg Nuclear Physics Institute, Gatchina, Leningrad region, 188300, Russia}
\newcommand{\pusan}{Pusan National University, Pusan 46241, Korea}
\newcommand{\riken}{RIKEN Nishina Center for Accelerator-Based Science, Wako, Saitama 351-0198, Japan}
\newcommand{\rikjrbrc}{RIKEN BNL Research Center, Brookhaven National Laboratory, Upton, New York 11973-5000, USA}
\newcommand{\rikkyo}{Physics Department, Rikkyo University, 3-34-1 Nishi-Ikebukuro, Toshima, Tokyo 171-8501, Japan}
\newcommand{\saispbstu}{Saint Petersburg State Polytechnic University, St.~Petersburg, 195251 Russia}
\newcommand{\seoulnat}{Department of Physics and Astronomy, Seoul National University, Seoul 151-742, Korea}
\newcommand{\stonybrkc}{Chemistry Department, Stony Brook University, SUNY, Stony Brook, New York 11794-3400, USA}
\newcommand{\stonycrkp}{Department of Physics and Astronomy, Stony Brook University, SUNY, Stony Brook, New York 11794-3800, USA}
\newcommand{\tenn}{University of Tennessee, Knoxville, Tennessee 37996, USA}
\newcommand{\titech}{Department of Physics, Tokyo Institute of Technology, Oh-okayama, Meguro, Tokyo 152-8551, Japan}
\newcommand{\tsukuba}{Tomonaga Center for the History of the Universe, University of Tsukuba, Tsukuba, Ibaraki 305, Japan}
\newcommand{\vandy}{Vanderbilt University, Nashville, Tennessee 37235, USA}
\newcommand{\weizmann}{Weizmann Institute, Rehovot 76100, Israel}
\newcommand{\wigner}{Institute for Particle and Nuclear Physics, Wigner Research Centre for Physics, Hungarian Academy of Sciences (Wigner RCP, RMKI) H-1525 Budapest 114, POBox 49, Budapest, Hungary}
\newcommand{\yonsei}{Yonsei University, IPAP, Seoul 120-749, Korea}
\newcommand{\zagreb}{Department of Physics, Faculty of Science, University of Zagreb, Bijeni\v{c}ka c.~32 HR-10002 Zagreb, Croatia}
\affiliation{\abilene}
\affiliation{\augie}
\affiliation{\banaras}
\affiliation{\barc}
\affiliation{\baruch}
\affiliation{\bnlcoll}
\affiliation{\bnlphys}
\affiliation{\caucr}
\affiliation{\charlesczech}
\affiliation{\cns}
\affiliation{\colorado}
\affiliation{\columbia}
\affiliation{\czechtech}
\affiliation{\debrecen}
\affiliation{\elte}
\affiliation{\eszterhazy}
\affiliation{\ewha}
\affiliation{\famu}
\affiliation{\fsu}
\affiliation{\gsu}
\affiliation{\hanyang}
\affiliation{\hiroshima}
\affiliation{\howard}
\affiliation{\ihepprot}
\affiliation{\illuiuc}
\affiliation{\inrras}
\affiliation{\instpasczech}
\affiliation{\isu}
\affiliation{\jaea}
\affiliation{\jeonbuk}
\affiliation{\jyvaskyla}
\affiliation{\kek}
\affiliation{\korea}
\affiliation{\kurchatov}
\affiliation{\kyoto}
\affiliation{\lahorelums}
\affiliation{\lawllnl}
\affiliation{\losalamos}
\affiliation{\lund}
\affiliation{\maryland}
\affiliation{\mass}
\affiliation{\michigan}
\affiliation{\muhlenberg}
\affiliation{\myongji}
\affiliation{\nara}
\affiliation{\natmephi}
\affiliation{\newmex}
\affiliation{\nmsu}
\affiliation{\northcg}
\affiliation{\ohio}
\affiliation{\ornl}
\affiliation{\orsay}
\affiliation{\pnpi}
\affiliation{\pusan}
\affiliation{\riken}
\affiliation{\rikjrbrc}
\affiliation{\rikkyo}
\affiliation{\saispbstu}
\affiliation{\seoulnat}
\affiliation{\stonybrkc}
\affiliation{\stonycrkp}
\affiliation{\tenn}
\affiliation{\titech}
\affiliation{\tsukuba}
\affiliation{\vandy}
\affiliation{\weizmann}
\affiliation{\wigner}
\affiliation{\yonsei}
\affiliation{\zagreb}
\author{U.~Acharya} \affiliation{\gsu} 
\author{A.~Adare} \affiliation{\colorado} 
\author{C.~Aidala} \affiliation{\michigan} 
\author{N.N.~Ajitanand} \altaffiliation{Deceased} \affiliation{\stonybrkc} 
\author{Y.~Akiba} \email[PHENIX Spokesperson: ]{akiba@rcf.rhic.bnl.gov} \affiliation{\riken} \affiliation{\rikjrbrc} 
\author{R.~Akimoto} \affiliation{\cns} 
\author{M.~Alfred} \affiliation{\howard} 
\author{N.~Apadula} \affiliation{\isu} \affiliation{\stonycrkp} 
\author{Y.~Aramaki} \affiliation{\riken} 
\author{H.~Asano} \affiliation{\kyoto} \affiliation{\riken} 
% ...
\collaboration{PHENIX Collaboration} \noaffiliation

\date{\today}


\begin{abstract}

The cross section of bottom quark-antiquark ($b\bar{b}$) production in 
$p$+$p$ collisions at $\sqrt{s}=510$ GeV is measured with the PHENIX 
detector at the Relativistic Heavy Ion Collider.  The results are based on 
the yield of high mass, like-sign muon pairs measured within the PHENIX 
muon arm acceptance ($1.2<|y|<2.2$).  The $b\bar{b}$ signal is extracted 
from like-sign dimuons by utilizing the unique properties of neutral $B$ 
meson oscillation.  We report a differential cross section of 
$d\sigma_{b\bar{b}\rightarrow \mu^\pm\mu^\pm}/dy = 0.16 \pm 
0.01~(\mbox{stat}) \pm 0.02~(\mbox{syst}) \pm 0.02~(\mbox{global})$ nb for 
like-sign muons in the rapidity and $p_T$ ranges $1.2<|y|<2.2$ and 
$p_T>1$~GeV/$c$, and dimuon mass of 5--10~GeV/$c^2$. The extrapolated total 
cross section at this energy for $b\bar{b}$ production is $13.1 \pm 
0.6~(\mbox{stat}) \pm 1.5~(\mbox{syst}) \pm 2.7~(\mbox{global})~\mu$b. The 
total cross section is compared to a perturbative quantum chromodynamics 
calculation and is consistent within uncertainties. The azimuthal opening 
angle between muon pairs from $b\bar{b}$ decays and their $p_T$ 
distributions are compared to distributions generated using {\sc ps 
pythia6}, which includes next-to-leading order processes.  The azimuthal 
correlations and pair $p_T$ distribution are not very well described by 
{\sc pythia} calculations, but are still consistent within uncertainties.  
Flavor creation and flavor excitation subprocesses are favored over gluon 
splitting.

\end{abstract}


\maketitle


\section{Introduction}
\label{Sec:intro}

The bottom-quark production in hadron-hadron collisions is an important 
test of perturbative quantum chromodynamics (pQCD) calculations. Because 
of its large mass, $m_b \gg \Lambda_{QCD}$, the $b$-quark production 
cross section can be reliably calculated by including next-to-leading 
order (NLO) processes, especially at high center of mass 
energies~\cite{Mangano1992295}.  The measurement of the \bb production 
cross section over a wide range of colliding energies in hadron-hadron 
collisions provides a meaningful test of pQCD theory calculations and a 
baseline measurement for studying modifications of heavy quark 
production in heavy ion collisions.

Cross section measurements for bottom production in hadron-hadron 
collision experiments have been made from lower energy fixed-target 
experiments~\cite{PhysRevLett.82.41,PhysRevLett.74.3118,PhysRevD.73.052005} 
($\sqrt{s} < 45$~GeV) up to collider energies ($\sqrt{s} > 100$~GeV). It 
was found that pQCD predictions match experimental results well at 
energies greater than $\sqrt{s} = 
1$~TeV~\cite{PhysRevD.71.032001,Aaij:2010gn,Aaij:2013noa,Aad:2012jga,ALICE_bb,Khachatryan:2011mk,Chatrchyan:2011pw,Chatrchyan:2012hw}, 
but less so at lower energies. Results at the wide range of collision energies
of the Relativistic Heavy Ion Collider explore an important 
gap between the low-energy fixed-target and TeV-energy regimes.

Without displaced vertex $b$-tagging capability at PHENIX, $b$-quark 
production has been studied using unlike-sign dileptons from heavy quark 
decays~\cite{Adare:2014iwg}. The PHENIX and STAR collaborations have 
previously measured the bottom cross section in \pp collisions at 
$\sqrt{s} =200$~GeV using electron-hadron 
correlations~\cite{PhysRevLett.103.082002,Aggarwal:2010xp} and using 
dilepton invariant mass and momentum 
distributions~\cite{physRevC.96.064901,Adare2009313,PhysRevD.99.072003}.

Like-sign dimuons have previously been used to investigate the phenomenon 
of neutral $B$ meson oscillations in $e^+e^-$ collisions by 
the CLEO Collaboration~\cite{PhysRevLett.58.183}, 
the ARGUS Collaboration~\cite{Albrecht1987245}, 
the ALEPH Collaboration~\cite{Buskulic1994441}, and 
in $p+\bar{p}$ collisions by the UA1 Collaboration~\cite{Albajar1987247}.
In this measurement, we use the yield of like-sign dimuons along with the 
properties of neutral $B$ meson oscillation to determine the \bb cross 
section. The correlated like-sign pairs at high mass (5--10 GeV/$c^2$) are 
dominated by the semileptonic decay of open bottom pairs and the other 
correlated sources (i.e. dijets or punch-through hadrons) amount to less 
than 10\%, and therefore provide a clean probe to study the \bb production.

\begin{figure}
\centering
\includegraphics[width=1\linewidth]{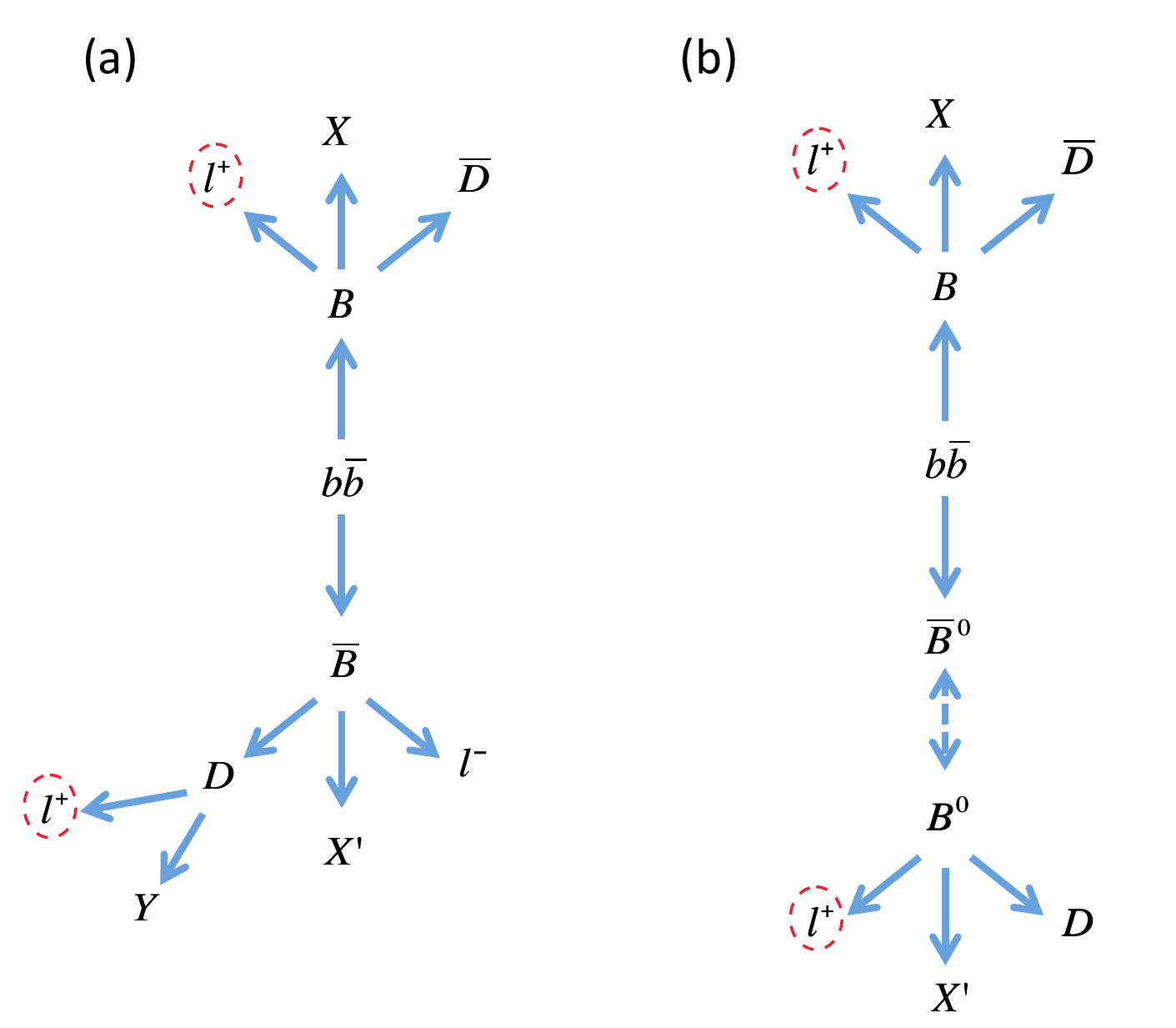}
\caption{\label{fig:decays} 
Example diagrams of lepton pair sources. (a) Like-sign primary-secondary 
or unlike-sign primary-primary dileptons from $B$ decay chain. (b) 
Primary-primary dileptons from neutral $B$ meson oscillation.
}
\end{figure}

In the Standard Model, neutral $B$ meson oscillation is a result of 
higher order weak interactions that transform a neutral $B$ meson into 
its antiparticle: $B^{0}\rightarrow\bar{B}^{0}$ because the flavor 
eigenstates differ from the physical mass eigenstates of the 
meson-antimeson system~\cite{Glashow:1961tr,Abe:1999ds}.  In the absence 
of oscillation as shown in Fig.~\ref{fig:decays}(a), primary-primary 
decays, where the lepton's direct parent is the $B$ meson, can only 
produce unlike-sign lepton pairs. For example $b\rightarrow 
\bar{B}(B^-,\bar{B}^0,\bar{B}^0_{s},..)\rightarrow l^-$ and 
$\bar{b}\rightarrow B(B^+,B^0,B^0_{s},..)\rightarrow l^+$ while like-sign 
lepton pairs can result from a mixture of primary and secondary decays 
(decay chain).

However, if oscillation occurs, as is the case for neutral $B$ mesons 
($B^0_d$ and $B^0_s$), the $\bar{B}^0$ meson can spontaneously change 
into a $B^0$ meson as shown in Fig.~\ref{fig:decays}(b). Unless 
otherwise noted, we denote $B$($\bar{B}$) as a generic admixture of 
bottom (antibottom) hadrons with production ratios, from weak decays 
($i.e.$ $Z\rightarrow b\bar{b}$) of: $B^+(B^-)=40.4\pm0.9\%$, 
$B^0(\bar{B}^0) = 40.4\pm0.9\%$, $B^0_s(\bar{B}^0_s) = 10.3\pm0.9\%$, 
and $b(\bar{b}$)-baryon $= 8.9\pm 1.5\%$~\cite{Agashe:2014kda}. The 
$B_c$ production ratio is negligible (0.2\%) and less than the 
uncertainties associated with bottom hadrons listed above. The 
time-integrated probability for a neutral $B$ meson to oscillate before 
it decays is defined as
\begin{equation}
\chi_{d/s} = \frac{1}{2} \frac{(\Delta m /\Gamma)^2}{1+(\Delta m/\Gamma)^2} \; ,
\end{equation}
where $\Delta m$ is the mass difference between heavy and light mass 
eigenstates and $\Gamma$ is the decay rate of the weak eigenstates. These 
values are found to be $\chi_{d} \approx 0.1874\pm 0.0018$ and $\chi_{s} 
\approx 0.499311\pm 0.000007$ for the $B^0_d$ and $B^0_s$ mesons, 
respectively~\cite{Agashe:2014kda}. This process can result in a 
like-sign dilepton event from a primary-primary decay as sown in 
Fig.~\ref{fig:decays}(b).  Given the large branching ratio of the 
$B\rightarrow\mu$ decay channel ($\approx 10.99\%$)~\cite{Agashe:2014kda}, 
the like-sign dilepton from a primary-primary decay provides a unique 
opportunity for extracting the \bb cross section.

In this paper, we present measurements of \bb production cross section 
through the like-sign dimuon decays and the azimuthal opening angle 
between the muon pair and their \pt distributions in \pp collisions at 
$\sqrt{s} = 510$~GeV at forward ($1.2<\!y<\!2.2$) and backward 
($-2.2\!<y\!<-1.2$) rapidities. The azimuthal opening angle and pair \pt 
distributions are compared to distributions generated using {\sc pythia6} 
with parton-shower ({\sc ps}) model~\cite{PYTHIA6}.
The model approximates the correction to all higher orders (almost 
next-to-leading-log) for \bb production, which includes 
flavor creation, flavor excitation, and gluon splitting. The extrapolated 
total cross section, using {\sc ps pythia6}~\cite{PYTHIA6} and 
{\sc pythia8}~\cite{SJOSTRAND2015159}, and MC$@$NLO~\cite{Frixione:2002ik} 
calculations, is also presented and compared to pQCD calculation.

The paper is organized as follows: The PHENIX apparatus is described in 
Sec.~\ref{sec:apparatus}. The data samples used for this analysis and the 
analysis procedure are presented in Sec.~\ref{sec:dataAna}. The results 
are presented and discussed in Sec.~\ref{sec:results}. The summary and 
conclusions are presented in Sec.~\ref{sec:summary}

\section{Experimental Setup}
\label{sec:apparatus}

\begin{figure}[htp!]
\centering
\includegraphics[width=0.99\linewidth,trim={0 10 0 413},clip]{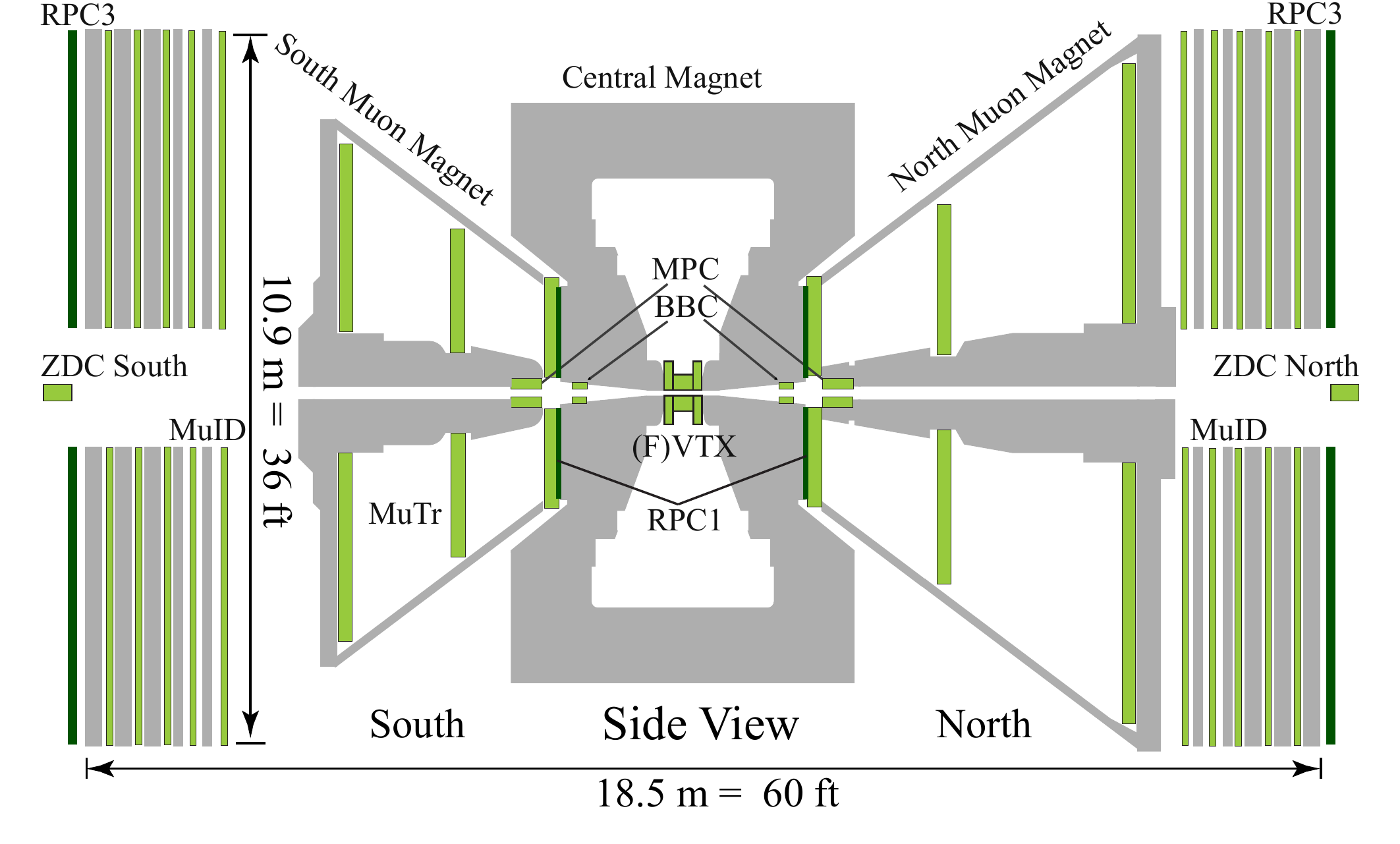}
\caption{\label{fig:Detector} 
A side view of the PHENIX detector, concentrating on the muon arm 
instrumentation.}
\end{figure}

A complete description of the PHENIX detector can be found in 
Ref.~\cite{Adcox:2003p2584}. We briefly describe here only the detector 
subsystems used in these measurements.  The relevant systems, which are 
shown in Fig.~\ref{fig:Detector}, include the PHENIX muon spectrometers 
covering forward and backward rapidities and the full azimuth. Each muon 
spectrometer comprises a hadronic absorber, a magnet, a muon tracker 
(MuTr), and a muon identifier (MuID). The absorbers comprise layers 
of copper, iron, and stainless steel and have about 7.2 interactions lengths. 
Following the absorber in each muon arm is the MuTr, which comprises three 
stations of cathode strip chambers in a radial magnetic field with an 
integrated bending power of 0.8~T$\cdot$m. The MuID comprises five 
alternating steel absorbers and Iarocci tubes. The composite momentum 
resolution, $\delta p/p$, of particles in the analyzed momentum range is 
about 5\%, independent of momentum and dominated by multiple scattering. 
Muon candidates are identified by reconstructed tracks in the muon 
spectrometers.

Another detector system relevant to this analysis is the beam-beam counter 
(BBC), consisting of two arrays of 64~\v{C}erenkov counters, located on both 
sides of the interaction point and covering the pseudorapidity range 
$3.1<|\eta|<3.9$. The BBC system was used to measure the \pp collision 
vertex position along the beam axis ($z_{\rm vtx}$), with 2 cm resolution, 
and initial collision time. It was also used to measure the beam luminosity 
and form a minimum bias trigger (MB). The MB trigger requires at least one 
hit in each BBC on the sides of the interaction point.

\section{Data Analysis}
\label{sec:dataAna}


\subsection{Data set and quality cuts}
\label{subsec:muid}

The data set for this analysis is collected by PHENIX during the 2013 \pp 
run at $\sqrt{s} = 510$~GeV. Events, in coincidence with the MB trigger, 
containing a muon pair within the acceptance of the spectrometer are 
selected by the level-1 dimuon trigger (MuIDLL1-2D) requiring that at 
least two tracks penetrate through the MuID to its last two layers. After 
applying a vertex cut of $|z_{\rm vtx}| < 30$ cm and extensive quality 
assurance checks, the data remaining correspond to $3.02\times10^{12}$ MB 
events or to an integrated luminosity of 94.4~pb$^{-1}$.

A set of cuts was used to select good muon candidates and improve the 
signal-to-background ratio. Hits in the MuTr are used to make MuTr tracks 
and hits in the MuID are used to make MuID roads. The MuTr track is 
required to have more than 9 hits out of the maximum possible of 16 while 
the MuID road is required to have more than 6 hits out of the maximum 
possible of 10. Additional $\chi^2$ cut is applied on MuTr track that is 
calculated from the difference between the measured hit positions of the 
track and the subsequent fit. MuTr tracks are then projected to the MuID 
at the first MuID gap and matched to MuID roads by applying cuts on 
maximum position and angle differences.

Muon candidates are required to have a minimum $p_T$ greater than 
1~GeV/$c$.  This cut improves the sample quality by reducing background 
from pions and kaons. A minimum of 3.0~GeV/$c$ is applied to single muon 
momentum along the beam axis, $p_z$, which is reconstructed and 
energy-loss corrected at the collision vertex, corresponding to the 
momentum cut effectively imposed by the absorbers. Muon candidates are 
further restricted to the rapidity range of $-2.2<y<-1.2$ for the south 
muon arm and $1.2 < y < 2.2$ for the north muon arm. Additionally, a cut 
on the $\chi^2$ of the fit of the two muon tracks to the common vertex of 
the two candidate tracks near the interaction point is applied.

\subsection{Detector acceptance and reconstruction efficiency}
\label{subsect:acc_eff}

The product of the acceptance and reconstruction efficiency ($A\epsilon$) 
is determined using Monte Carlo (MC) simulation. The $A\epsilon$ is 
defined by the number of dimuons reconstructed in the muon spectrometers 
with respect to the number of dimuons generated in the same kinematic 
region. The kinematic distributions of {\sc pythia}\footnote{We used 
{\sc pythia6} (ver 6.421), with parton distribution functions given by 
CTEQ6LL. The following parameters were modified: MSEL = 0, MSUB(86) = 1, 
PARP(91) = 2.1, MSTP(51) = 10041, MDME(858,1) = 0, MDME(859,1) = 1, 
MDME(860,1) = 0, and Tune A.}~\cite{Field:2005sa} generated $p_T$, 
rapidity, and \bb mass shape were used as input into a full 
PHENIX {\sc geant4} simulation~\cite{AGOSTINELLI2003250}.

The $p_T$ and rapidity distributions were tuned such that the 
reconstructed distributions match those of 2013 data. Variations within 
the uncertainties of data are taken as systematic uncertainty.

The detector response in the simulation is tuned to a set of characteristics 
(dead and hot channel maps, gains, noise, etc.) that describes the 
performance of each detector subsystem. The simulated vertex distribution is 
also tuned to match that of the 2013 data. The simulated events are further 
embedded with real data to account for the effects of detector noise and 
other background tracks, and then are reconstructed in the same manner as 
the real data. A final cross check was done on $J/\psi$ invariant yield 
after $A\epsilon$ correction, which matched very well within statistical 
uncertainties in all $p_T$ and rapidity bins~\cite{PhysRevD.101.052006}. 

\begin{figure}
\centering
\includegraphics[width=1\linewidth]{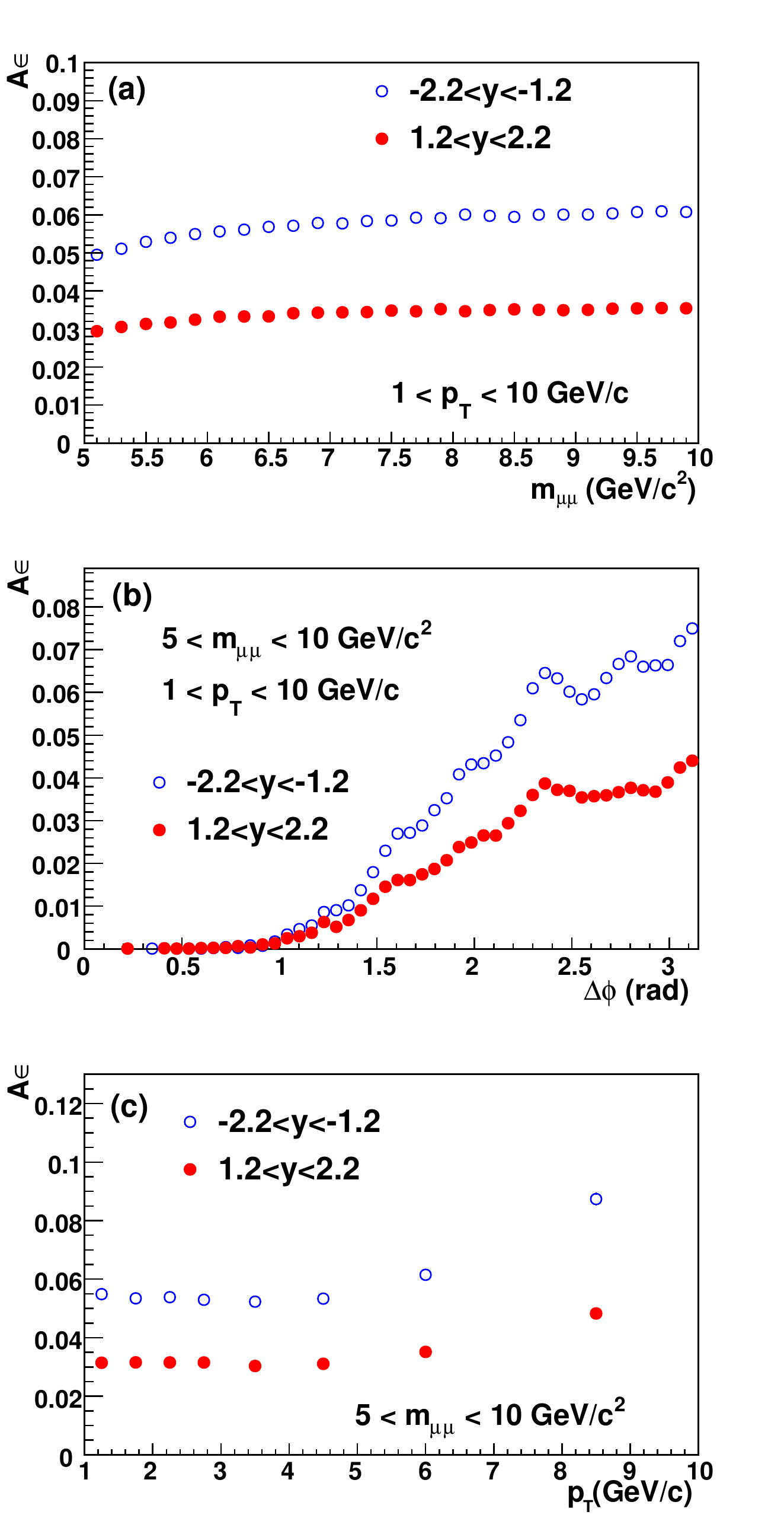}
\caption{\label{fig:AccEff} 
$A\epsilon$ as a function of (a) invariant mass for like-sign dimuons, 
(b) dimuon azimuthal opening angle, and (c) dimuon \pt.  Shown are 
the weighted averages of $\mu^+\mu^+$ and $\mu^-\mu^-$ distributions.
}
\end{figure}

Figure~\ref{fig:AccEff} shows the $A\epsilon$ as a function of (a) dimuon 
mass $m_{\mu\mu}$, (b) dimuon opening angle $\delta\Phi$, and (c) dimuon 
$p_T$.  The relative difference in $A\epsilon$ between the two spectrometers 
is due to different detection efficiencies of the MuTr and MuID systems and 
different amounts of absorber material.

\subsection{Raw yield extraction}

We measure like-sign dimuons in the same muon arm that have an invariant 
mass between 5 and 10~GeV/$c^2$. In this mass range, the correlated pairs in 
the dimuon spectrum are dominated by the semileptonic decay of open bottom 
pairs either from the primary-secondary decay chain as shown in 
Fig.~\ref{fig:decays}(a) or from the primary-primary pairs from neutral $B$ 
meson oscillation as shown in Fig.~\ref{fig:decays}(b). Dileptons from the 
Drell-Yan process and quarkonia decays can only yield unlike-sign pairs. $D$ 
mesons can produce like-sign pairs through their decay chain. For example, 
$c\rightarrow D^+ \rightarrow \mu^+ + anything$ and the other open charm 
decays as $\bar{c} \rightarrow D^- \rightarrow K^+ + anything \rightarrow 
\mu^+\nu_{\mu}$. However, in the mass range of interest the like-sign pairs 
from $D$ mesons are negligible. The contribution from neutral $D$ meson 
oscillation to the like-sign signal is expected to be very small because the 
oscillation probability is 
$\mathcal{O}(<10^{-2})$~\cite{PhysRevLett.110.101802}; therefore, it is not 
included.

\subsubsection{Correlated background}

Additional contribution to the correlated pairs could originate from 
correlated sources such as dijets or punch-through hadrons. Hadrons 
(particularly $\pi^\pm$ and $K^\pm$) can punch through to the last gap of 
the MuID or decay to muons creating a background to the correlated 
like-sign signal. These contributions are estimated using MC simulation 
by determining the \pt-dependent survival probability that a hadron will 
traverse the muon arm detectors and applying it to {\sc pythia} generated 
dihadron pairs to get the yield expected at the back of the muon arm 
detectors. $\pi^\pm$ and $K^\pm$ are generated with {\sc 
pythia}\footnote{Non-default parameters used in Multiparton Interaction 
(MPI)``Tune-A" {\sc ps pythia6} simulation for hadron and jet 
production. The following parameters were modified:MSEL = 1, PMAS(5,1) = 
4.1, PYTUNE 100, and PARP(90) = 
0.25}~\cite{Field:2005sa,PhysRevD.99.072003,PhysRevD.84.012006} and then 
run through the PHENIX detector simulation chain to determine a 
$p_T$-dependent probability that the hadrons penetrate the last gap of 
the MuID.

\begin{figure}
\centering
\includegraphics[width=1\linewidth]{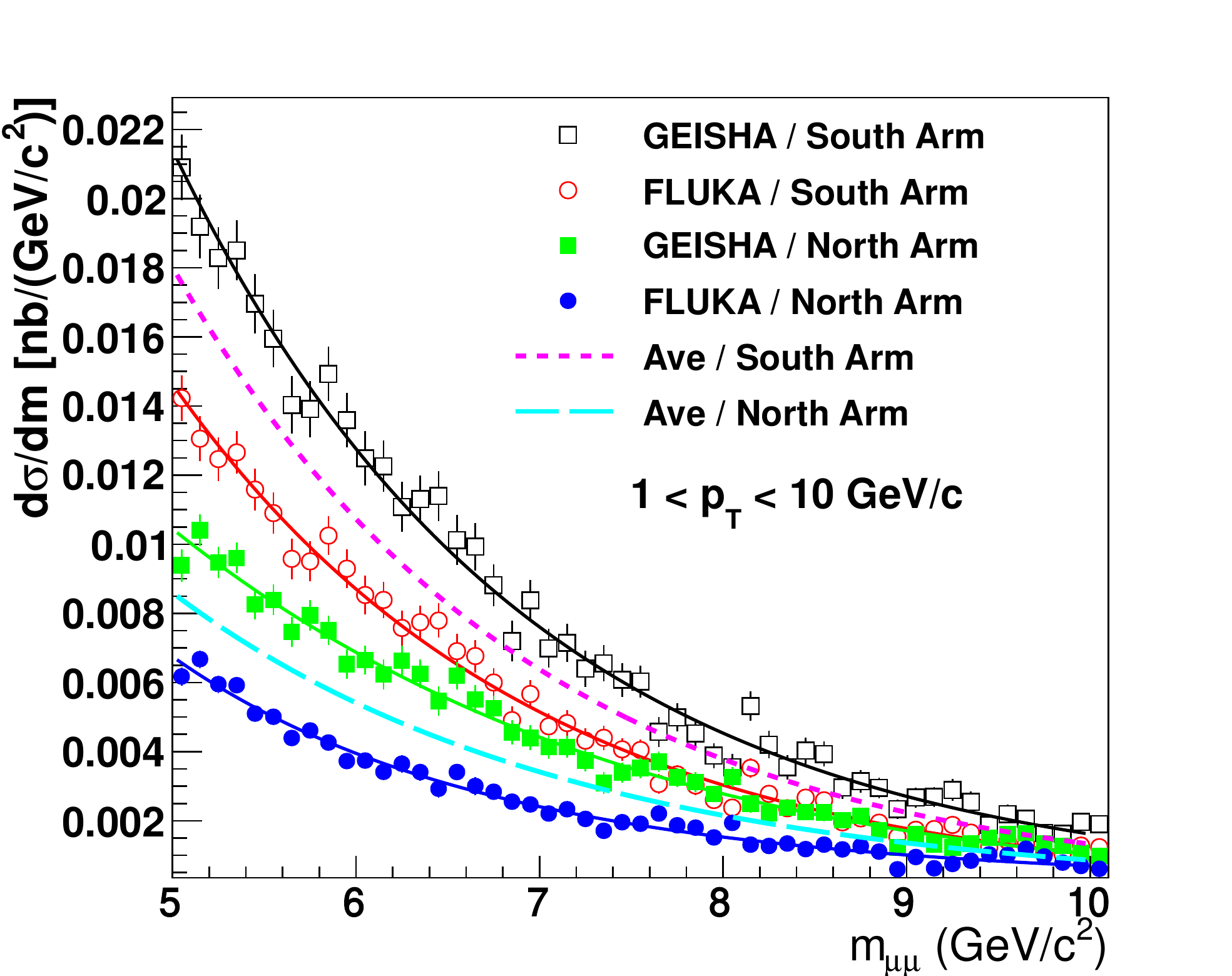}
\caption{\label{fig:Jet_sim} 
Like-sign invariant mass distribution from jet background simulation in 
the north and south arms. The solid lines are fits to the data with an 
exponential function between 5 and 10 GeV/$c^2$ while the dashed lines 
represent the averages of the resulting fits.
}
\end{figure}

To get a better estimate of the survival probability, the hadron 
simulation is run using two different hadron interaction packages for 
{\sc geant:} {\sc fluka} and 
{\sc geisha}~\cite{Brun:1994aa,PhysRevC.86.024909}. 
Figure~\ref{fig:Jet_sim} shows the simulated invariant mass spectra from 
irreducible background are fitted with an exponential function of the 
form exp($a + b \times m + c \times m^2$) between 5 and 10 GeV/$c^2$, 
where $m$ is the invariant mass and $a$, $b$ and $c$ are fit parameters. 
The average of the indicated results from {\sc geisha} and {\sc fluka} 
is used to subtract the hadronic background from like-sign pairs while 
the difference is considered as a systematic uncertainty.

The invariant mass distribution for like-sign pairs is then constructed 
from {\sc pythia} generated dihadron pairs within the same event and from 
mixed events, with each entry weighted by the survival probability. 
Event-mixing procedure is discussed in the next section. Just as with 
data, the correlated like-sign signal is obtained by subtracting the 
mixed event spectrum from the like-sign spectrum, providing the 
correlated like-sign signal due to dijets or punch-through hadrons.  The 
sum of $\pi$ and $K$ correlated like-sign signals is weighted based on 
their $p_T$-dependent cross sections~\cite{PhysRevD.98.032007, 
NuclPhysB.335.261}.
 
Fake like-sign pairs due to charge misidentification and like-sign pairs 
from Drell-Yan process or quarkonia decays and muon-decayed or 
punch-through hadrons were also studied and found to be negligible.

\subsubsection{Uncorrelated background}

The uncorrelated pair contribution is estimated using event mixing 
technique~\cite{Crochet2002564}, where like-sign pairs are constructed by 
pairing muons in the current event with those of the same sign and same 
arm in previous events of z-vertex position within 2 cm. The mixed event 
pairs ($N_{++}^{BG}$ and $N_{--}^{BG}$) form the uncorrelated background 
spectrum which is normalized to the the foreground ($N_{++}^{FG}$ and 
$N_{--}^{FG}$) using a normalization factor 
($\sqrt{N_{++}^{FG}N_{--}^{FG}}/\sqrt{N_{++}^{BG}N_{--}^{BG}}\;$). The 
normalization factor requires that the integrated counts from event 
mixing equal those from the like-sign in the low mass region where the 
correlated pairs are expected to be negligible~\cite{Crochet2002564}. The 
normalized like-sign pairs from event mixing are given as:
\begin{equation}
 \label{eq:Nbckd}
 N_{\pm\pm}^{BG} = \left(  N_{++}^{BG} + N_{--}^{BG} \right) \frac{\sqrt{N_{++}^{FG}N_{--}^{FG}}}{\sqrt{N_{++}^{BG}N_{--}^{BG}}}.
\end{equation}

However, the specific range where the signal of interest is negligible is 
not well known, and the average of normalization factors over five mass 
ranges (0.6--2.6~GeV/$c^2$, 1.0--2.0~GeV/$c^2$, 1.6--3.2~GeV/$c^2$, 
2.6--3.8~GeV/$c^2$, and 0.6--4.2~GeV/$c^2$) is used.  The correlated 
like-sign signal ($N_{\pm\pm}^{\rm cor}$) is then isolated by subtracting 
the mixed-event spectrum ($N_{\pm\pm}^{BG}$) from the ``foreground'' 
like-sign pairs ($N_{\pm\pm}^{FG}$) according to the following,

\begin{equation}
\label{eq:Ncorr}
N_{\pm\pm}^{\rm cor} = N_{\pm\pm}^{FG}-N_{\pm\pm}^{BG} .
\end{equation}

To further improve the normalization process, the \bb invariant mass 
distribution shape from {\sc ps pythia6} simulation is utilized. This is 
done by normalizing the integral of the {\sc ps pythia6} distribution to 
the result of Eq.~(\ref{eq:Ncorr}), over the signal mass range 
5--10~GeV/$c^2$. The integral of the normalized \bb mass distribution is 
then subtracted from the background distribution in Eq.~(\ref{eq:Nbckd}) 
for each of the background ranges and the normalization factor is 
recalculated. The second step is then repeated until the value of the 
mixed-events normalization factor converges.

\begin{figure}
\includegraphics[width=1\linewidth]{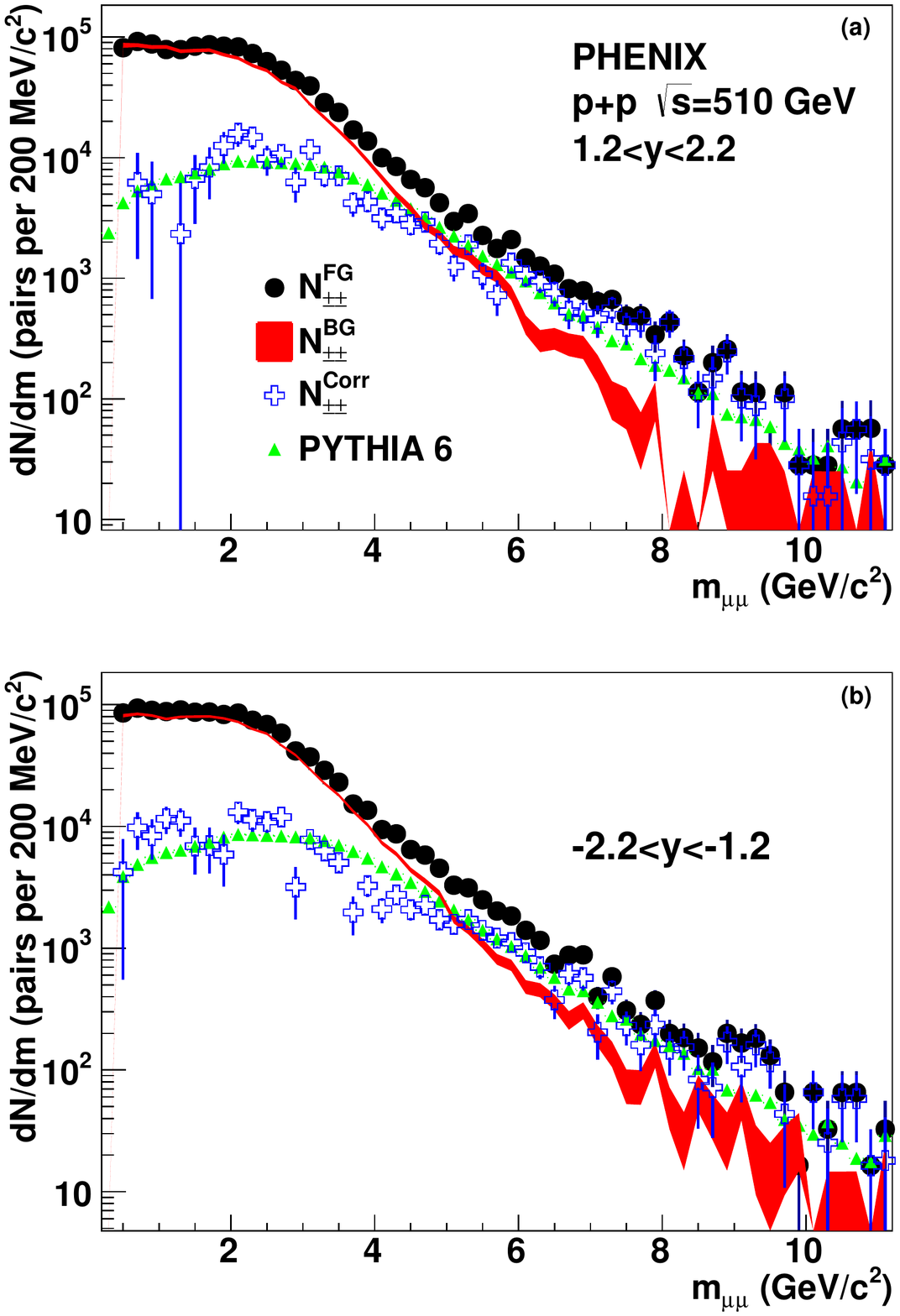}
\caption{\label{fig:spectra1}  
Invariant mass spectra for like-sign pairs from the same event 
($N_{\pm\pm}^{FG}$, solid black points), event-mixing ($N_{\pm\pm}^{BG}$, 
red band), and the difference between the two ($N_{\pm\pm}^{\rm cor}$, empty 
blue pluses) for the (a) north arm and (b) south arm. These distributions 
are corrected with $A\epsilon$. The solid green triangles show {\sc 
pythia} \bb shape.
}
\end{figure}

Figure~\ref{fig:spectra1} shows the resulting distributions of 
$N_{\pm\pm}^{FG}$, $N_{\pm\pm}^{BG}$ and $N_{\pm\pm}^{\rm cor}$ as a 
function of the invariant mass of the pairs. These distributions are 
corrected with $A\epsilon$.
To extract the \bb distribution as a function of the azimuthal opening 
angle between muon pairs (\dphi) and their \pt, the normalization factors 
obtained previously are used to normalize \dphi and \pt mixed event 
distributions, which are then subtracted from \dphi and \pt foreground 
distributions, respectively.

\subsection{Systematic uncertainties}

Table~\ref{tab:sysUncer} summarizes the systematic uncertainties.
Evaluated as standard deviations, they are divided into three categories
based upon the effect each source has on the measured results:

\begin{description} 

\item[Type-A] 
Point-to-point uncorrelated uncertainties that allow the data points to 
move independently with respect to one another and are added in 
quadrature with statistical uncertainties; however, no systematic 
uncertainties of this type are associated with this measurement.

\item[Type-B]
Point-to-point correlated uncertainties which allow the data points to 
move coherently within the quoted range to some degree. These systematic 
uncertainties include a 4\% uncertainty from MuID tube efficiency and an 
8.2\% (2.8\%) from MuTr overall efficiency for the north (south) arm. The 
systematic uncertainty associated with $A\epsilon$ includes the 
uncertainties on the input \pt and rapidity distributions which are 
extracted by varying these distributions over the range of the 
statistical uncertainty of the data, yielding 4.4\% (5.0\%) for the north 
(south) arm. To be consistent with the real data analysis, a trigger 
emulator was used to match the MuIDLL1-2D trigger for the data. The 
efficiency of the trigger emulator was studied by comparing the dimuon 
mass spectrum requiring the dimuon passes the trigger emulator to the 
dimuon mass spectrum requiring the dimuon passes the MuIDLL1-2D trigger, 
which resulted in a 1.5\% (2\%) uncertainty for the north (south) arm. 
Additional 11.2\% (8.8\%) systematic effect for the north (south) arm was 
also considered to account for the azimuthal angle distribution 
difference between data and simulation.

The source of systematic uncertainty in signal extraction is the 
normalization of mixed events which could come from the choice of the 
different normalization ranges in the mixed events or \bb shape from 
{\sc pythia} used to guide the signal extraction. A 1.9\% uncertainty on 
the mixed events normalization was observed from using each of the five 
normalization windows by itself as well as the different combinations of 
these normalization windows. {\sc pythia} \bb shape is the sum of three 
subprocesses: flavor creation, flavor excitation and gluon splitting. A 
maximum variation of 3.1\% on the extracted signal was observed from 
choosing each of the subprocesses by itself as the source of \bb shape. 
Added in quadrature, they result in a 3.6\% uncertainty on signal 
extraction.

The systematic uncertainty associated with correlated backgrounds could 
come from the input \pt distribution, differences between {\sc geisha} and 
{\sc fluka}, and differences between {\sc geant3} and {\sc geant4}. 
{\sc pythia} \pt distributions of $\pi^\pm$ and $K^\pm$ were compared 
separately to fits of UA1 data~\cite{PhysRevD.98.032007, 
NuclPhysB.335.261} and an overall difference of 18\% was observed. 
Differences of up to 30\% and 20\% between {\sc fluka} and {\sc geisha}, 
see Fig.~\ref{fig:Jet_sim}, were observed in the north and south arms, 
respectively. Additional 15\% was considered to account for the 
difference between {\sc geant3} and {\sc geant4}. Added in quadrature, 
all three sources give an overall effect on the hadronic background of 
39\% (31\%) for the north (south) arm for the mass and $\Delta\phi$ 
distributions. For \pt distribution, a \pt-dependent correction was used 
for the effect on the input \pt spectra and the other two sources gave 
an overall effect on the hadronic background of 34\% (25\%) for the 
north (south) arm. To extract the systematic uncertainty associated with 
the cross section (or invariant yields) for all distributions (mass, 
$\Delta\phi$ and \pt ), the hadronic background was varied between the 
limits listed above which resulted in an overall systematic of 5.1\% 
(4.5\%) for north (south) arm.

The Type-B systematic uncertainties are added in quadrature and amount to 
16.0\% (12.8\%) for the north (south) arm. They are shown as shaded bands 
on the associated data points.

\item[Type-C]
An overall (global) normalization uncertainty of 10\% was assigned for 
the BBC cross section and efficiency 
uncertainties~\cite{PhysRevLett.91.241803} which allows the data points 
to move together by a common multiplicative factor.

\end{description} 

\begin{table}[ht!]
\caption{\label{tab:sysUncer} 
Systematic uncertainties associated with the differential cross section 
calculation in the north (south) arm.}
\begin{ruledtabular} \begin{tabular}{ccc}
Type & Origin & North (South)\\
\hline
B & MuID hit efficiency & 4.0\% (4.0\%)\\
B & MuTr hit efficiency & 8.2\% (2.8\%)\\
B & \accEff \pt and $y$ input distributions & 4.4\% (5.0\%)\\
B & \accEff trigger emulator & 1.5\% (2.0\%)\\
B & \accEff $\phi$ distribution & 11.2\% (8.8\%)\\
B & Signal extraction & 3.6\% (3.6\%)\\
B & Correlated background & 5.1\% (4.5\%)\\
B & Quadratic sum & 16.4\% (12.8)\%\\
C & MB trigger efficiency & 10\%\\
\end{tabular} \end{ruledtabular}
\end{table}

\section{Results and Discussion}
\label{sec:results}

\subsection{Differential cross section}

The differential yield and cross section of $B$ meson pairs decaying into 
like-sign dimuons as a function of mass are calculated according to the 
following relations,
\begin{equation}
\label{eq:invYield}
\frac{d^2N}{dydm}= \frac{1}{\Delta y\Delta m} \frac{N_{\mu\mu}}{A\epsilon(y,m)}\frac{\epsilon_{\rm BBC}^{\rm MB}}{\epsilon^{\rm BBC}N_{\rm MB}}  \; ,
\end{equation}
\begin{equation}
\label{eq:diff_xs}
\frac{d^2\sigma}{dydm}= \frac{d^2N}{dydm}\frac{\sigma_{\rm BBC}^{pp}}{\epsilon_{\rm BBC}^{\rm MB}}  \; ,
\end{equation}
where $N_{\mu\mu}/A\epsilon(y,m)$ is the yield of like-sign dimuons from 
$B$ meson decay normalized by $A\epsilon(y,m)$ in $y$ and $m$ bin with 
$\Delta y$ and $\Delta m$ widths, respectively. The yields of the north 
and south arms are calculated independently and are consistent within 
statistical uncertainties; therefore, the weighted 
average~\cite{EPJC.74.3004} is used in the differential yield 
calculation. $\sigma_{\rm BBC}^{pp} = 32.5 
\pm 3.2$ mb is the cross section as seen for the BBC in \pp collisions at 
$\sqrt{s} = 510$~GeV, which is determined from the van der Meer scan 
technique~\cite{PhysRevLett.106.062001}. $\epsilon_{\rm BBC}^{\rm MB} = 0.53 \pm 0.02$ is the fraction of inelastic \pp collisions recorded by the BBC~\cite{PhysRevD.79.012003}. $\epsilon^{\rm BBC}=0.91 \pm 0.04$ 
is the efficiency of the MB trigger for events containing a hard 
scattering~\cite{PhysRevD.101.052006}. $N_{\rm MB}$ is the number of MB 
events.

\begin{figure}[htp!]
\includegraphics[width=1\linewidth]{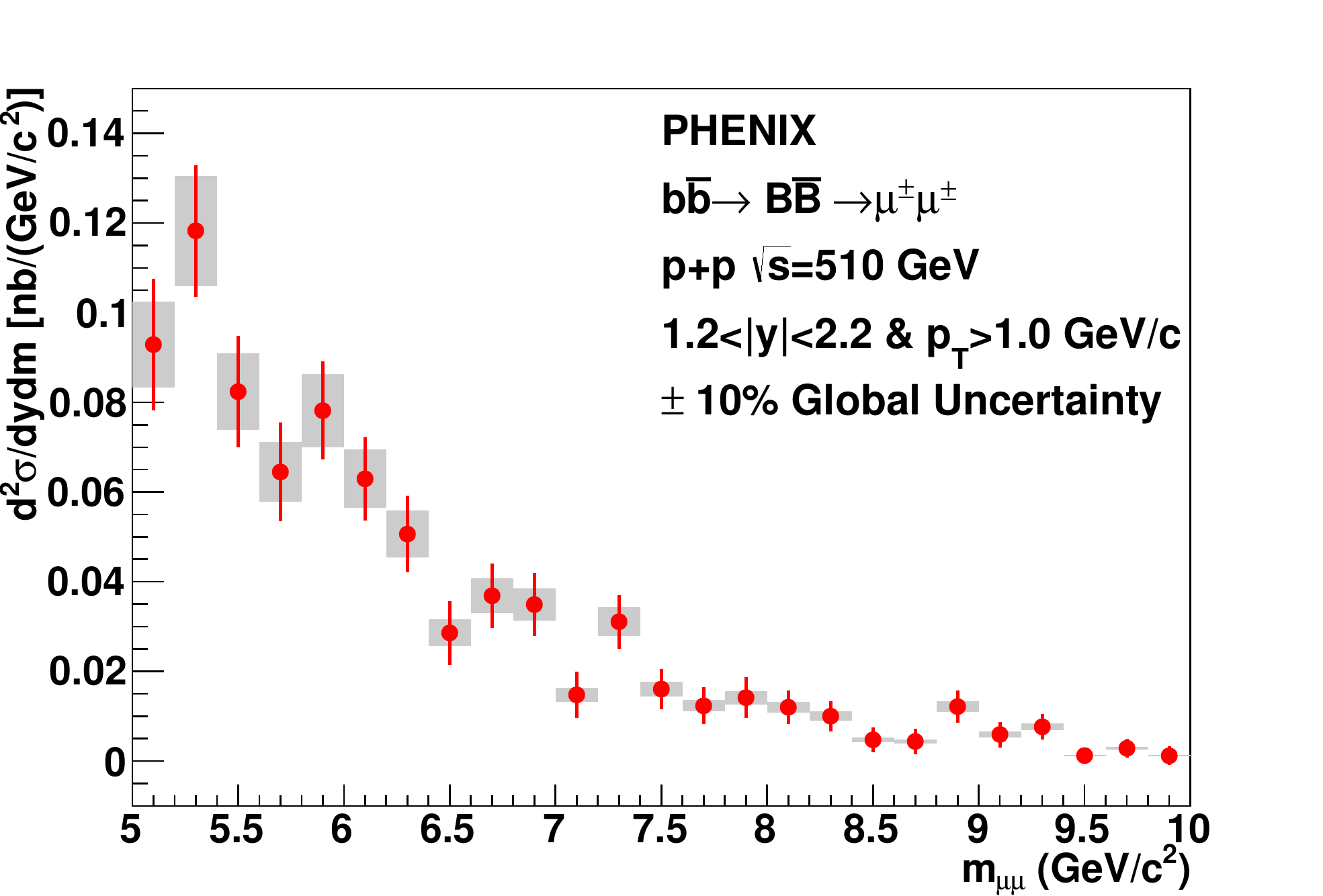}
\caption{\label{fig:diffXsec_BBbar_osc} 
Differential cross section of like-sign dimuons from $B$ meson decay. The 
error bars represent the statistical uncertainties, and the gray shaded 
bands represent the quadratic sum of type-B systematic uncertainties.
}
\end{figure}

The differential cross section of like-sign dimuons from $B$ meson decay 
is shown in Fig.~\ref{fig:diffXsec_BBbar_osc}.
The gray shaded bands represent the weighted average of the quadratic sum 
of type-B systematic uncertainties of the north and south arms, 
$\approx$10.1\%. The average is weighted based on the statistical 
uncertainties of each arm. In addition to type-B systematic 
uncertainties, we have a 10\% global systematic uncertainty for BBC cross 
section and efficiencies~\cite{PhysRevLett.91.241803}.

The total cross section, $d\sigma_{b\bar{b}\rightarrow 
B\bar{B}\rightarrow\mu^\pm\mu^\pm }/dy$, within the mass range, $5 < 
m_{\mu^\pm\mu^\pm} < 10$ GeV/$c^2$, and rapidity and \pt ranges, 
$1.2<|y|<2.2$ and \pt $>$ 1~GeV/$c$, respectively, is extracted by 
integrating $d^2\sigma_{b\bar{b}\rightarrow 
B\bar{B}\rightarrow\mu^\pm\mu^\pm }/dydm$, which resulted 
$d\sigma_{b\bar{b}\rightarrow\mu^\pm\mu^\pm 
}/dy~(1.2<|y|<2.2,~p_T>1~\mbox{GeV}/c,~5<m_{\mu^\pm\mu^\pm}<10~\mbox{GeV}/c^2) 
= 0.16\pm0.01 ~\mbox{(stat)}\pm0.02~\mbox{(type-B syst)} \pm 
0.02~(\mbox{global syst})$ nb.

To obtain the differential cross section of all $B$ meson pairs that 
decay into dimuons, regardless of the muon pair charge, the differential 
cross section of like-sign dimuons from $B$ meson decay is scaled by the 
ratio of the total number of all $B$ meson pairs that decay into dimuons, 
regardless of their sign, to those of like-sign. For clarification 
purposes, the process is divided into two separate steps defined by two 
variables $\alpha(m)$ and $\beta$, both of which depend on the signal 
from like-sign dimuons due to oscillation.

\begin{figure}
\includegraphics[width=1\linewidth]{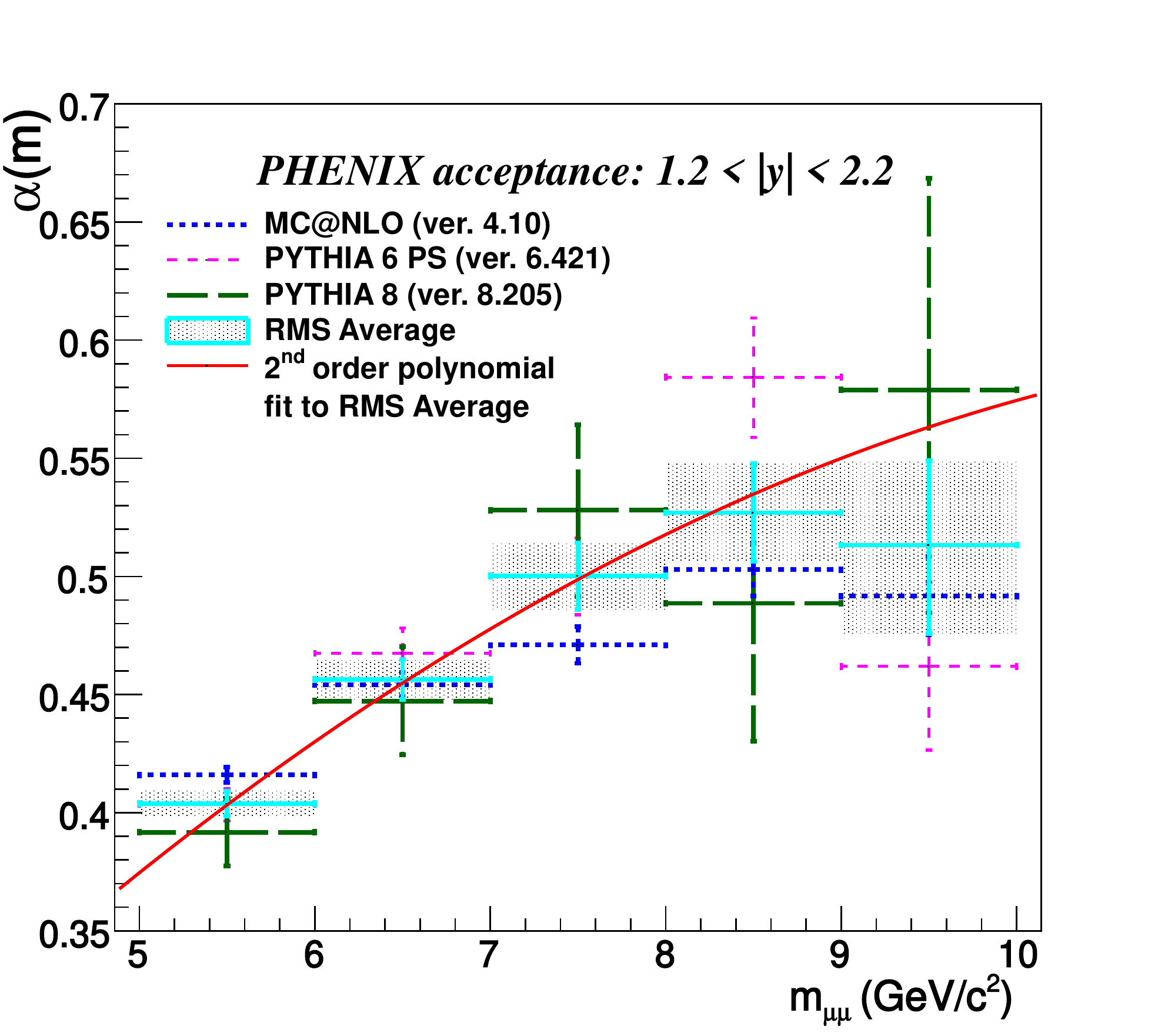}
\caption{\label{fig:alpha} 
Fraction of like-sign dimuons from neutral $B$ meson oscillation 
($\alpha(m)$) from {\sc mc@nlo} (blue points) , {\sc ps pythia6} 
(magenta points) and {\sc pythia8} (green points) within the PHENIX 
muon-arms acceptance. Cyan data points are the RMS average of the three model 
calculations. The shaded boxes are the associated errors based on the 
three model calculations. The red curve is a second-order polynomial fit 
to the RMS data points.
}
\includegraphics[width=1\linewidth]{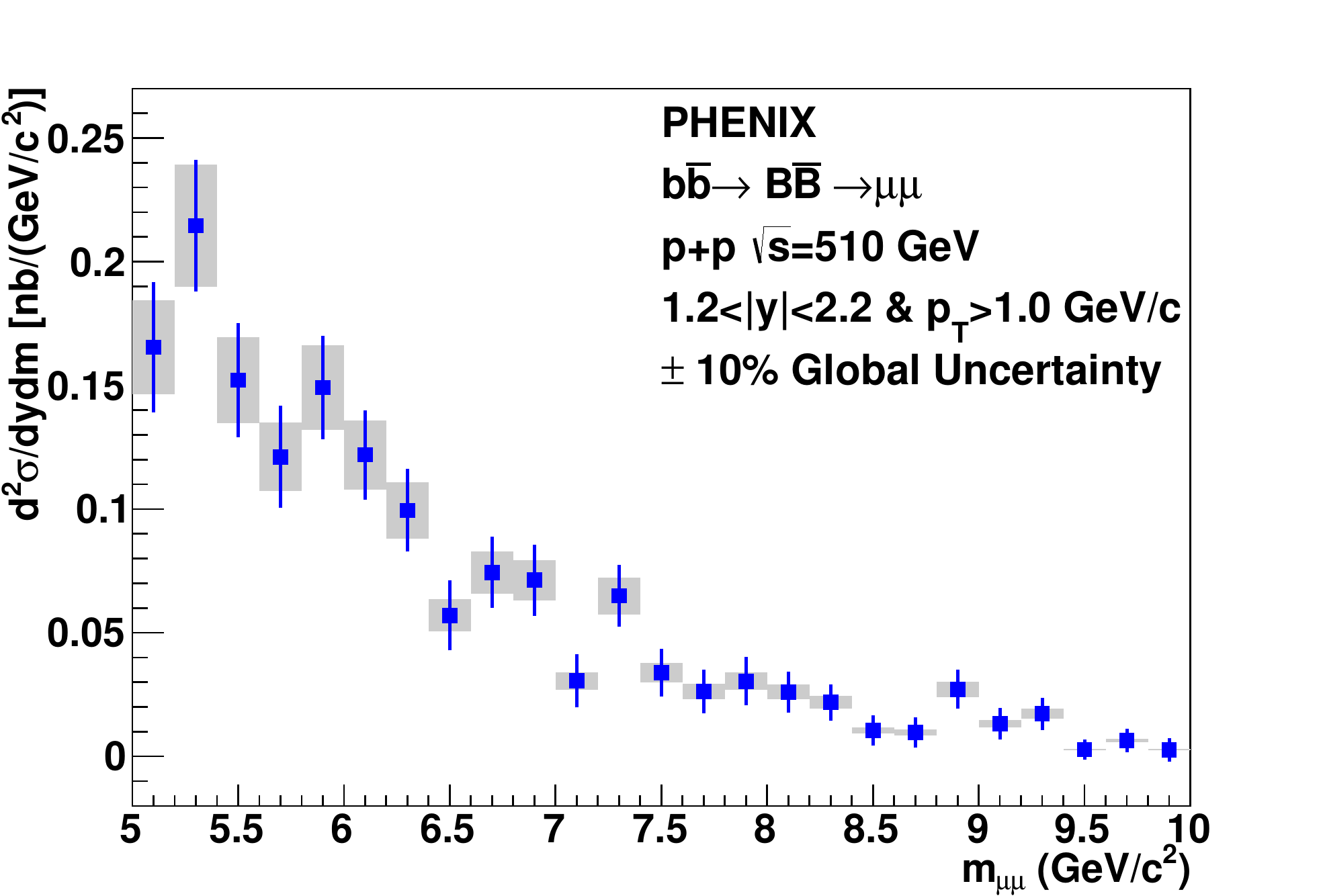}
\caption{\label{fig:diffXsec_BBbar} 
Differential cross section of all dimuons from $B$ meson decay. The 
error bars represent the statistical uncertainties, and the gray shaded 
band represents the quadratic sum of type-B systematic uncertainties.
}
\end{figure}

The ratio of like-sign dimuons at mass $m$ and from primary-primary 
decays due to $B^0$ oscillation to like-sign muon pairs resulting from 
primary-primary or a mixture of primary-secondary decays is defined as:
\begin{equation}\label{EQ:ALPHA}
\alpha(m)=\frac{b\bar{b}\rightarrow B\bar{B} \rightarrow\mu^{\pm}\mu^{\pm} \mbox{ (osc)} }{b\bar{b}\rightarrow B\bar{B}\rightarrow \mu^{\pm}\mu^{\pm} },
\end{equation}
which is calculated in the mass range $5 < m < 10~\mbox{GeV}/c^2$ at 
$1.2<|y|<2.2$ and \pt $>$ 1 GeV/$c$ and extrapolates the correlated 
like-sign signal to an open bottom signal from oscillation, 
$N_{\pm\pm}^{osc}$.  The $\alpha(m)$ is obtained using open bottom 
events from three model calculations: {\sc mc@nlo} (ver 4.10), 
{\sc ps pythia6} (ver 6.421) and {\sc pythia8} (ver 8.205) as shown in 
Fig.~\ref{fig:alpha}. The red line is a second-order polynomial fit 
with $\chi^2/ndf$ of 3.8/4.  The shaded boxes represent the uncertainty 
based on the three model calculations.
 
\begin{figure*}
\begin{minipage}{0.75\linewidth}
\includegraphics[width=0.99\linewidth]{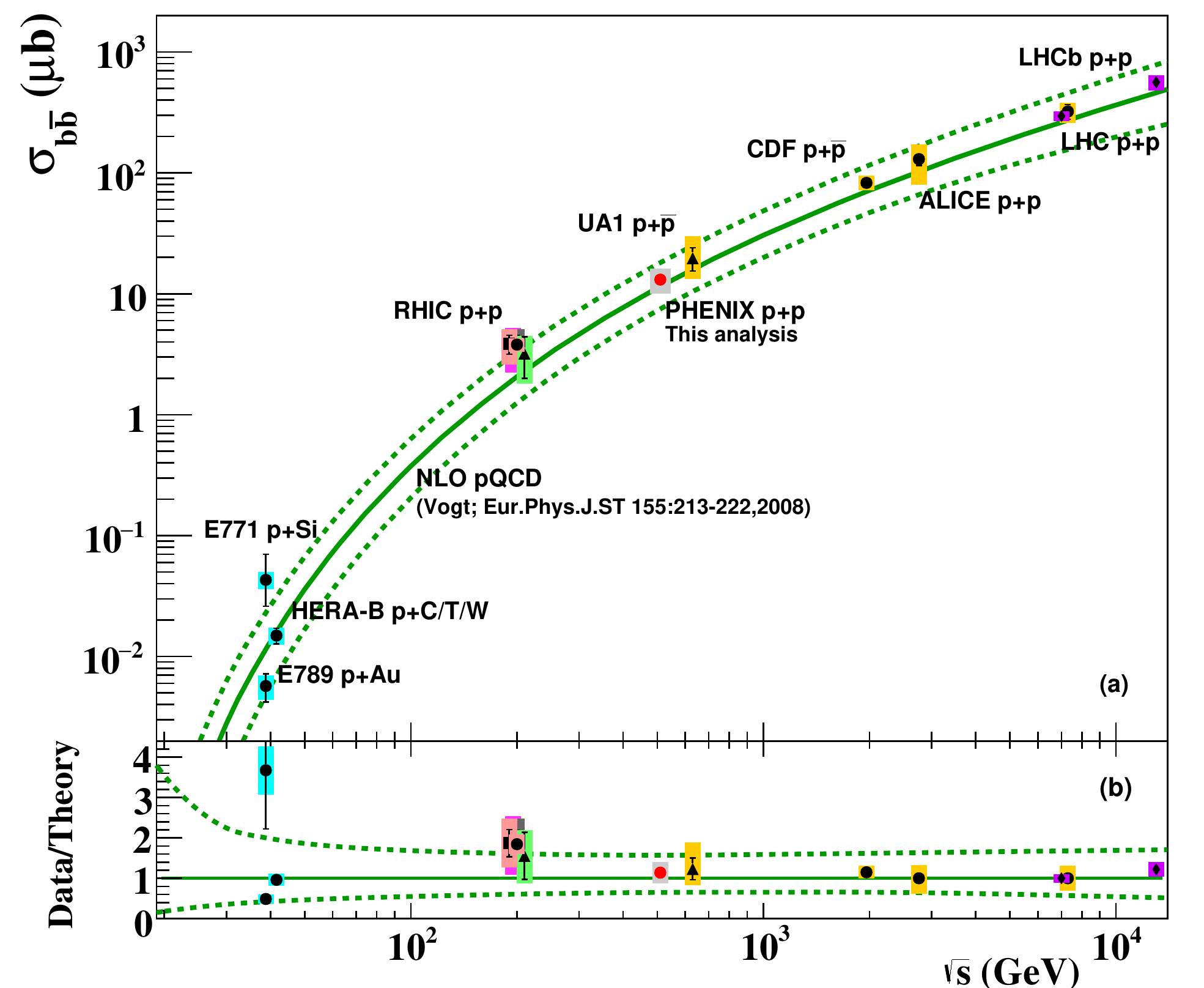}
\end{minipage}
\begin{minipage}{0.23\linewidth}
\caption{\label{fig:totXsec}
(a) Bottom cross section, $\sigma_{b\bar{b}}$ as a function of 
$\sqrt{s}$. The curves are NLO pQCD calculation~\cite{Vogt} with the 
dashed lines being error bands obtained by varying the renormalization 
scale, factorization scale and bottom quark mass. (b) Ratio of data to 
NLO pQCD calculation.
}
\end{minipage}
\end{figure*}

\begin{figure*}[ht]
\begin{minipage}{0.48\linewidth}
\includegraphics[width=0.99\linewidth]{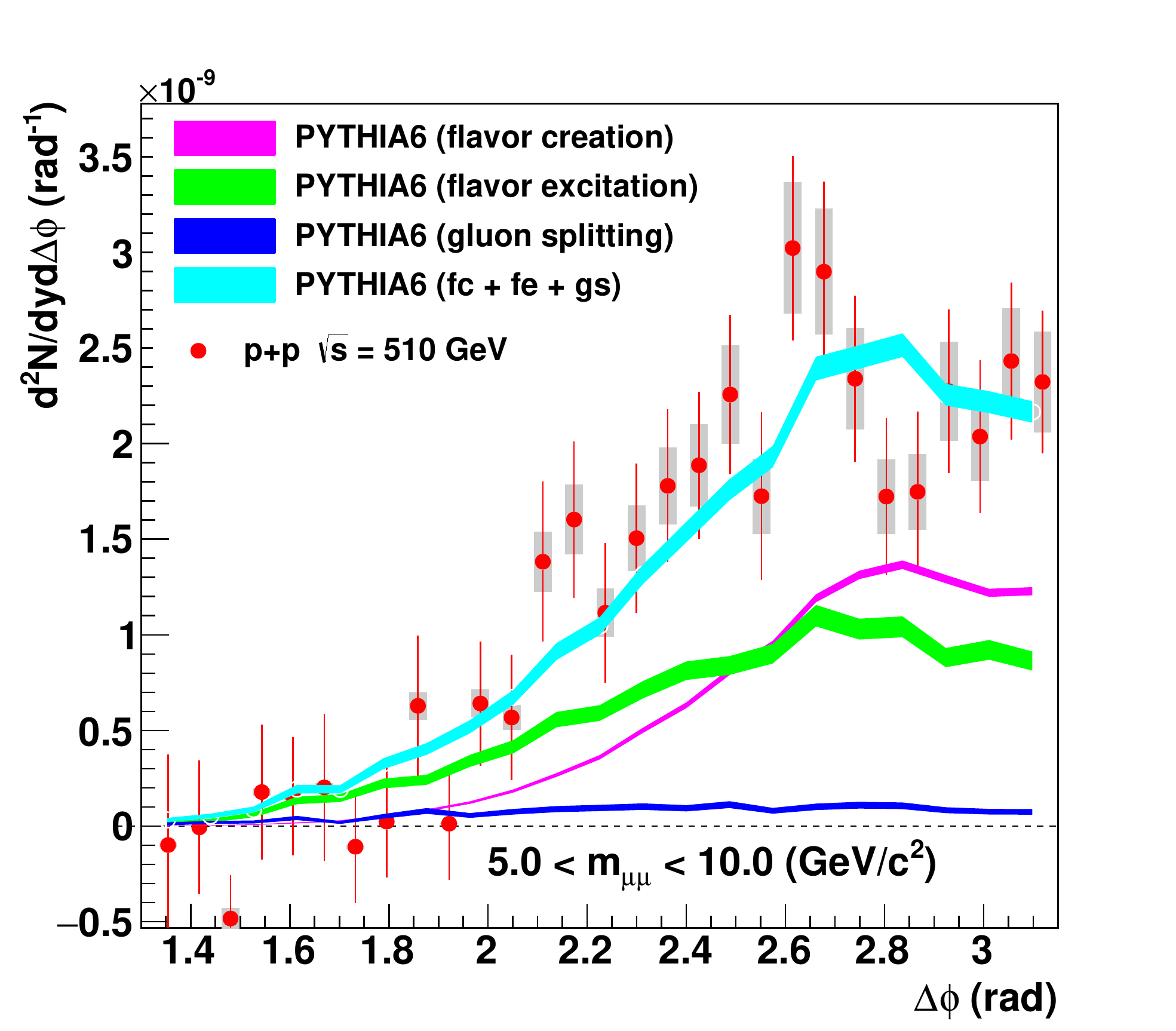}
\caption{\label{fig:dndphi_theory1} 
Like-sign $\mu\mu$ yield as a function of the azimuthal opening angle. 
The data are compared to the distributions calculated with 
{\sc ps pythia6}. The model calculations are normalized to the data. For 
{\sc ps pythia6} the $\mu\mu$ pair yield is broken down into contributions 
from flavor creation, flavor excitation, and gluon splitting.
}
\end{minipage}
\hspace{0.4cm}
\begin{minipage}{0.48\linewidth}
\includegraphics[width=0.99\linewidth]{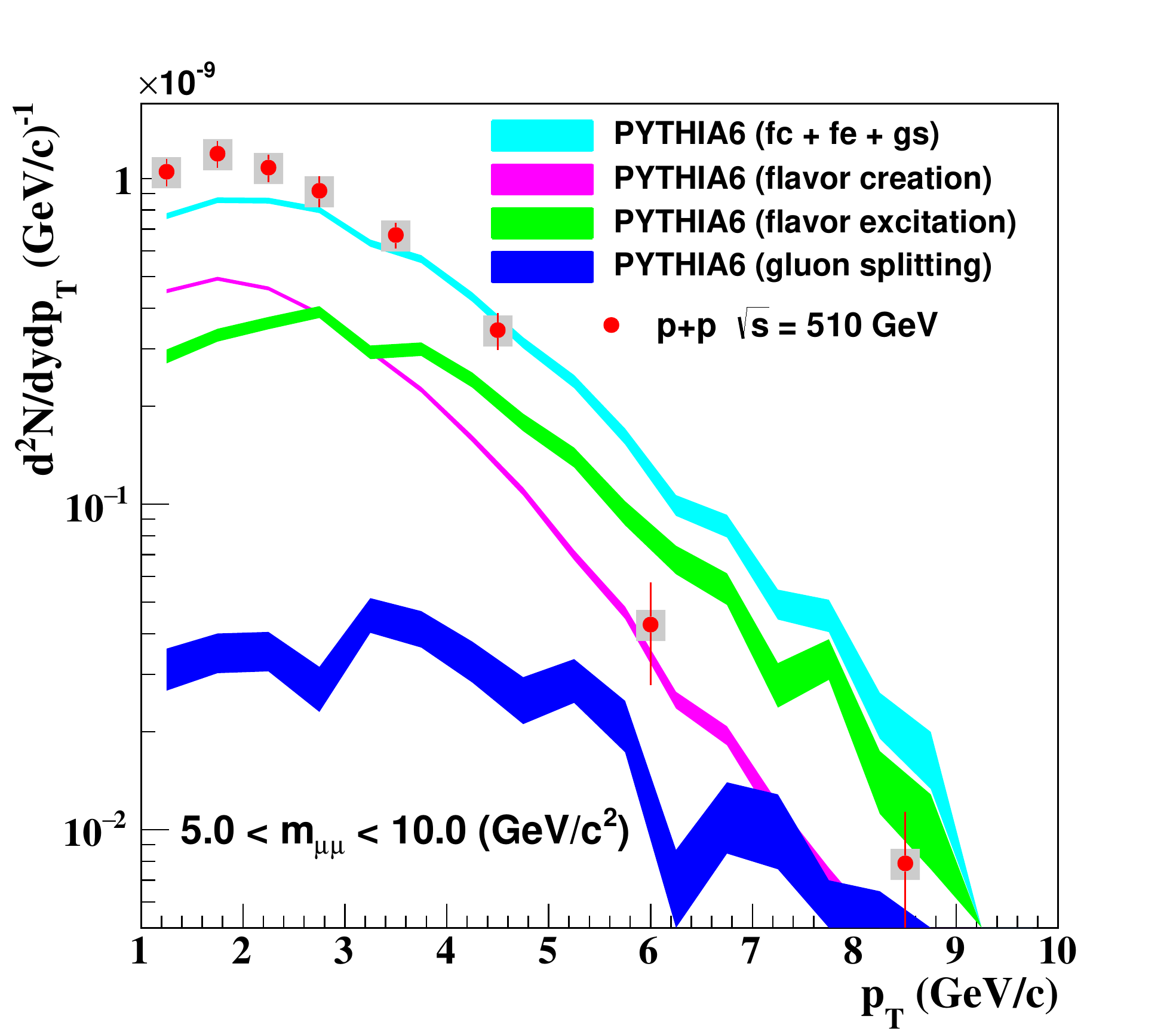}
\caption{\label{fig:dndpt_theory} 
Like-sign $\mu\mu$ yield as a function of the pair \pt. The data are 
compared to the distributions calculated with {\sc ps pythia6}. The 
model calculations are normalized to the data. For {\sc ps pythia6} the 
$\mu\mu$ pair yield is broken down into contributions from flavor 
creation, flavor excitation, and gluon splitting.}
\end{minipage}
\end{figure*}

$\beta$ is the ratio of primary-primary like-sign dimuons due to $B^0$ 
oscillation to all $B$ meson pairs that decay into primary-primary 
dimuons with all possible muon charge pairs ($++, --$ and $+-$). 
$\beta$ converts the number of muon pairs from oscillation into all $B$ 
meson pairs and is defined as: \begin{equation}\label{EQ:BETA} \beta = 
\frac{b\bar{b}\rightarrow B\bar{B} \rightarrow \mu^{\pm}\mu^{\pm} \mbox{ 
(osc)}}{b\bar{b}\rightarrow B\bar{B} \rightarrow \mu\mu} . \end{equation} 
The value of $\beta$ is $0.22\pm0.01$ which is the calculated RMS value 
from the three model simulations described above. The error of $\beta$ is 
the standard deviation of the three model calculations which represents 
the model-dependent uncertainty.

The differential cross section of all $B$ meson pairs that decay into a 
primary-primary dimuon, regardless of the muon pair charge, is then 
calculated as follows:

\begin{equation}\label{BB_eqn}
\frac{d^2\sigma_{b\bar{b}\rightarrow B\bar{B}\rightarrow\mu\mu }}{dydm}=\frac{\alpha(m)}{\beta}\frac{d^2\sigma_{b\bar{b}\rightarrow B\bar{B}\rightarrow\mu^{\pm}\mu^{\pm}}}{dydm}.
\end{equation}
Figure~\ref{fig:diffXsec_BBbar} shows the differential cross section of 
all $B$ meson pairs that decay into a primary-primary dimuon.
Additional type-B systematic uncertainties associated with this 
measurement due to $\alpha(m)$ and $\beta$ and amount to 1.9\% and 
4.5\%, respectively, are included. This brings the type-B systematic 
uncertainties on $d^2\sigma_{b\bar{b}\rightarrow 
B\bar{B}\rightarrow\mu\mu }/dydm$ to 11.2\%.

The total cross section, $d\sigma_{b\bar{b}\rightarrow 
B\bar{B}\rightarrow\mu\mu }/dy$, within the mass range, $5.0 < 
m_{\mu\mu} < 10.$ GeV/$c^2$, and rapidity and \pt ranges, $1.2<|y|<2.2$ 
and \pt $>$ 1~GeV/$c$, respectively, is extracted by integrating 
$d^2\sigma_{b\bar{b}\rightarrow B\bar{B}\rightarrow\mu\mu }/dydm$, which 
resulted $d\sigma_{b\bar{b}\rightarrow\mu\mu 
}/dy~(1.2<|y|<2.2,~p_T>1~\mbox{GeV}/c,~5<m_{\mu\mu}<10~\mbox{GeV}/c^2) = 
0.31\pm0.01 ~\mbox{(stat)}\pm0.04~\mbox{(type-B syst)} \pm 
0.03~(\mbox{global syst})$ nb.

\subsection{Total cross section}

To extrapolate from the \bb differential cross section in the muon decay 
channel within the acceptance of muon arms to a total \bb cross section, 
the differential cross section is scaled by the ratio of $B$ pairs that 
decay to dimuons within the measured region to those over the entire 
kinematic range. This method is similar to that found in 
Ref.~\cite{ZEUS}.

The total cross section, $\sigma_{b\bar{b}}$, is extrapolated and 
corrected for the semileptonic branching ratio in the following manner:

\begin{equation}
\sigma_{b\bar{b}} = \frac{d\sigma_{b\bar{b}\rightarrow\mu\mu}}{dy} \times  \frac{1}{scale}  \times  \frac{1}{(BR_{B\rightarrow\mu})^{2}} \; ,
\end{equation}
where $BR_{B\rightarrow\mu}$ is the branching ratio of $B$ to muon 
through the primary decay channel (=10.99$\%$), and $scale$, defined as:

\begin{equation}\label{eq_scale}
scale = \frac{B\bar{B}\rightarrow\mu\mu (1.2 < y < 2.2 ; p_T>1; 5<m_{\mu\mu}<10)}{B\bar{B}\rightarrow\mu\mu (all)} ,
\end{equation}
which is a factor used to convert from the visible kinematic region to full 
phase space. The $scale$ factor is determined from {\sc pythia} and {\sc 
mc@nlo} simulations. It is taken as the average value, 
$1.96\times10^{-3}$, of {\sc ps pythia6} (CTEQ6LL), {\sc ps pythia6} 
(CTEQ5M1), {\sc pythia8} (CTEQ6LL) and {\sc mc@nlo} (CTEQ5M) as listed 
in Table~\ref{tab:scale_sys}.

\begin{table}[ht]
\caption{\label{tab:scale_sys} 
Values of the scale factor as found using 
{\sc ps pythia6}~\protect\cite{PYTHIA6},
{\sc pythia8}~\protect\cite{SJOSTRAND2015159}, 
and MC$@$NLO~\protect\cite{Frixione:2002ik}.
}
\begin{ruledtabular} \begin{tabular}{lc}
Simulation & Scale Factor \\
\hline
{\sc pythia8} (CTEQ6LL) &  0.00210\\
{\sc ps pythia6} (CTEQ6LL) &  0.00207\\
{\sc ps pythia6} (CTEQ5M1) &  0.00255\\
{\sc mc@nlo} (CTEQ5M) & 0.00113\\
Average Value & 0.00196\\
\end{tabular} \end{ruledtabular}
\end{table}

The difference in the $scale$ factor due to the different models and 
parton distribution functions is considered to be a global type-C 
uncertainty, which amounts to 18.1\%. This results in a total cross 
section of $13.1 \pm 0.6~(\mbox{stat}) \pm 1.5~(\mbox{type-B syst}) \pm 
2.7~(\mbox{global syst})~\mu$b. Type-B systematic uncertainties are from 
the differential cross section while global uncertainties are the 
quadrature sum of type-C from the differential cross section and 
uncertainties arising from the extrapolation.

The $\sigma_{b\bar{b}}$ measured at $\sqrt{s} = 510$~GeV is shown in 
Fig.~\ref{fig:totXsec} and compared to measurements from other 
experiments~\cite{Adare2009313,PhysRevLett.82.41,PhysRevLett.74.3118,PhysRevD.73.052005,Albajar1991121,ALICE_bb}.  
The solid line is the cross section from NLO pQCD 
calculations~\cite{Vogt} and the dashed lines are error bands, and they 
are obtained by varying the renormalization scale, factorization scale 
and bottom quark mass. At $\sqrt{s} = 510$~GeV, the NLO pQCD calculation 
predicts $\sigma_{b\bar{b}} = 11.5^{+6.5}_{-3.9}~\mu$b, which is 
consistent with the extrapolated total cross section using the current 
dimuon analysis within uncertainties. Figure~\ref{fig:totXsec} also shows the ratio of data to theory.

\subsection{Azimuthal correlations and pair \pt}

The like-sign $\mu\mu$ pair yield from \bb decays is shown in 
Fig.~\ref{fig:dndphi_theory1} and Fig.~\ref{fig:dndpt_theory} as a function 
of $\Delta\phi$ and pair \pt, respectively. The spectra are compared to 
model calculations based on {\sc ps pythia6} that are normalized by fitting 
the subprocesses sum to the data~\cite{PhysRevD.99.072003}. The generated 
pairs are filtered with the same kinematic cuts that are applied in the data 
analysis.

The azimuthal opening angle distribution from {\sc ps pythia6} shows a 
similar pattern to that of the data, an increase until $\approx$2.6 rad 
and then drop, and it is consistent with the data with 
$\chi^2/ndf{\approx}27/28$, when considering the quadrature sum of the 
statistical and systematic uncertainties. The data show steeper \pt 
dependence than that of {\sc ps pythia6} but they are still consistent 
when considering the large statistical and systematic uncertainties. We 
note that flavor creation fits the data much better than any other 
subprocess with $\chi^2/ndf{\approx}8.4/7$. These results show similar 
behavior to that observed at 200 GeV~\cite{PhysRevD.99.072003} where the 
data favors a dominant mix of flavor creation and flavor excitation 
subprocesses over gluon splitting.

\section{Summary and Conclusion}
\label{sec:summary}

In summary, we presented first measurements of the differential cross 
section for dimuons from bottom quark-antiquark production in \pp collisions 
at $\sqrt{s}=510$ GeV, which we found to be: $d\sigma_{b\bar{b}\rightarrow 
\mu^\pm\mu^\pm}/dy = 0.16 \pm 0.01~(\mbox{stat}) \pm 0.02~(\mbox{syst}) \pm 
0.02~(\mbox{global})$ nb. The analysis technique is based on the yield of 
high-mass correlated like-sign dimuons from parity-violating decays of $B$ 
meson pairs at forward and backward rapidities. Using a model dependent 
extrapolation, the measured differential cross section is converted to a 
total cross section of $13.1 \pm 0.6~(\mbox{stat}) \pm 1.5~(\mbox{syst}) \pm 
2.7~(\mbox{global})~\mu$b. This extrapolated total cross section is 
consistent with NLO pQCD calculations within uncertainties. This agreement 
with NLO pQCD calculations at $\sqrt{s}=510$ GeV is better than what was 
observed at 200 GeV~\cite{PhysRevD.99.072003}, possibly indicating a better match 
with NLO pQCD calculations at higher energies.  However, the measurement at 
$\sqrt{s}=200$ GeV used the unlike-sign pairs method and could be impacted 
by the presence of Drell-Yan process and resonances.

The azimuthal opening angle between the muons from \bb decays and the pair 
\pt distributions are compared to distributions generated using 
{\sc ps pythia6}~\cite{PYTHIA6}, which includes NLO processes.  While the 
data tend to have a wider azimuthal distribution than {\sc ps pythia6} 
calculations and present a steeper \pt distribution, both are still 
consistent within uncertainties with {\sc ps pythia6}, where flavor creation 
and flavor excitation subprocesses are dominant. This is similar to what was 
observed at 200 GeV~\cite{PhysRevD.99.072003}.



\begin{acknowledgments}

We thank the staff of the Collider-Accelerator and Physics
Departments at Brookhaven National Laboratory and the staff of
the other PHENIX participating institutions for their vital
contributions.  We acknowledge support from the
Office of Nuclear Physics in the
Office of Science of the Department of Energy,
the National Science Foundation,
Abilene Christian University Research Council,
Research Foundation of SUNY, and
Dean of the College of Arts and Sciences, Vanderbilt University
(U.S.A),
Ministry of Education, Culture, Sports, Science, and Technology
and the Japan Society for the Promotion of Science (Japan),
Conselho Nacional de Desenvolvimento Cient\'{\i}fico e
Tecnol{\'o}gico and Funda\c c{\~a}o de Amparo {\`a} Pesquisa do
Estado de S{\~a}o Paulo (Brazil),
Natural Science Foundation of China (People's Republic of China),
Croatian Science Foundation and
Ministry of Science and Education (Croatia),
Ministry of Education, Youth and Sports (Czech Republic),
Centre National de la Recherche Scientifique, Commissariat
{\`a} l'{\'E}nergie Atomique, and Institut National de Physique
Nucl{\'e}aire et de Physique des Particules (France),
Bundesministerium f\"ur Bildung und Forschung, Deutscher Akademischer
Austausch Dienst, and Alexander von Humboldt Stiftung (Germany),
J. Bolyai Research Scholarship, EFOP, the New National Excellence
Program ({\'U}NKP), NKFIH, and OTKA (Hungary),
Department of Atomic Energy and Department of Science and Technology
(India),
Israel Science Foundation (Israel),
Basic Science Research and SRC(CENuM) Programs through NRF
funded by the Ministry of Education and the Ministry of
Science and ICT (Korea).
Physics Department, Lahore University of Management Sciences (Pakistan),
Ministry of Education and Science, Russian Academy of Sciences,
Federal Agency of Atomic Energy (Russia),
VR and Wallenberg Foundation (Sweden),
the U.S. Civilian Research and Development Foundation for the
Independent States of the Former Soviet Union,
the Hungarian American Enterprise Scholarship Fund,
the US-Hungarian Fulbright Foundation,
and the US-Israel Binational Science Foundation.

\end{acknowledgments}




%
 
\end{document}